\documentclass[%
 reprint,
 amsmath,amssymb,
 aps,
]{revtex4-2}

\usepackage{graphicx}
\usepackage{dcolumn}
\usepackage{bm}
\usepackage{hyperref}
\usepackage{slashed}
\usepackage{subfloat}
\usepackage{pgfplots,subfigure}
\usepackage{orcidlink}

\usepackage{geometry}
\definecolor{acsblue}{RGB}{17,76,139}

\begin{document}
\fontsize{7.7}{8.7}\selectfont
\preprint{APS/123-QED}

\title{Optics in spiral dislocation spacetime: Torsion as a geometric waveguide and frequency-filtering mechanism}

\author{Semra Gurtas Dogan\orcidlink{0000-0001-7345-3287}}
\email{semragurtasdogan@hakkari.edu.tr}
\affiliation{Department of Medical Imaging Techniques, Hakkari University, 30000, Hakkari, Türkiye}

\author{Omar Mustafa\orcidlink{0000-0001-6664-3859}}
\email{omar.mustafa@emu.edu.tr (Corr. Auth.)}
\affiliation{Department of Physics, Eastern Mediterranean University, 99628, G. Magusa, north Cyprus, Mersin 10 - Türkiye}

\author{Abdullah Guvendi\orcidlink{0000-0003-0564-9899}}
\email{abdullah.guvendi@erzurum.edu.tr}
\affiliation{Department of Basic Sciences, Erzurum Technical University, 25050, Erzurum, Türkiye}

\author{Hassan Hassanabadi\orcidlink{0000-0001-7487-6898}}
\email{hha1349@gmail.com }
\affiliation{Department of Physics, Faculty of Science, University of Hradec Králové, Rokitanského 62, 500 03 Hradec Králové, Czechia}

\begin{abstract}
{\fontsize{7.7}{8.7}\selectfont
\vspace{0.10cm}
\setlength{\parindent}{0pt}
We present an exact analytical investigation of null trajectories and scalar wave propagation in a $(2+1)$-dimensional spacetime containing a spiral dislocation-a topological defect characterized by torsion in the absence of curvature. For null rays, the torsion parameter $\beta$ modifies the affine structure, enforcing a finite turning radius $r_{\min} = \sqrt{b^2 - \beta^2}$ and inducing a torsion-mediated angular deflection that decreases monotonically with increasing $\beta$. The photon trajectory deviates from the curvature-induced lensing paradigm, exhibiting a purely topological exclusion zone around the defect core. In the wave regime, we recast the Helmholtz equation into a Schrödinger-like form and extract a spatially and spectrally dependent refractive index $n^2(r,k)$. This index asymptotically approaches unity at large distances, but diverges strongly and negatively near the dislocation core due to torsion-induced geometric terms. The resulting refractive index profile governs the transition from propagating to evanescent wave behavior, with low-frequency modes experiencing pronounced localization and suppression. Our findings reveal that torsion alone, absent any curvature, can act as a geometric regulator of both classical and quantum propagation, inducing effective anisotropy, frequency filtering, and confinement. This framework provides a rare exact realization of light-matter interaction in a torsion-dominated background, with potential applications in analog gravity systems and photonic metamaterials engineered to replicate non-Riemannian geometries.}
\end{abstract}


\maketitle

\section{Introduction}

\vspace{0.10cm}
\setlength{\parindent}{0pt}

Topological defects in spacetime, which emerge as exact solutions to Einstein’s field equations or as effective geometrical models of matter distributions, exhibit profound geometrical and physical implications with broad relevance to cosmology, high-energy physics, and condensed matter theory~\cite{kibble1976, vilenkin1994, mermin1979,torsion}. Among the various classes of defects, spacetime with spiral dislocation represents a screw-type distortion along a privileged spatial direction, extending the geometry of space to incorporate torsional effects~\cite{torsion,furtado1994, katanaev1992, puntigam1997}. This geometry is characterized by a non-trivial dislocation parameter \(\beta\), which introduces intrinsic torsion into the spacetime manifold and alters its global topological properties~\cite{torsion,furtado1994, desousa1990,puntigam1997,katanaev1992, kleinert1989, OA, HH, New-2, ZHA}. Physically, spiral dislocations serve as gravitational analogues of screw dislocations in crystalline media, where atomic planes are displaced helicoidally~\cite{katanaev1992, kleinert1989, OA, HH, ZHA}. In gravitational contexts, these configurations offer idealized, yet insightful models for line-like singularities that potentially arise in early universe scenarios or in analog gravity models describing topologically non-trivial materials~\cite{torsion,vilenkin1994, galtsov2012, Leonhardt2009, chen2010}.

\vspace{0.10cm}
\setlength{\parindent}{0pt}

The presence of a non-zero dislocation parameter \(\beta\) induces significant modifications in the spacetime causal structure and affects the geodesic dynamics. Specifically, photon trajectories acquire a torsional component that alters their propagation paths~\cite{guven,silva}, in contrast to the purely radial or angular motion found in torsion-free geometries. This leads to a variety of novel effects, including modified light deflection, direction-dependent lensing, and effective birefringence of electromagnetic waves~\cite{guven,silva}. Furthermore, the coupling of \(\beta\) to the conserved quantities arising from spacetime symmetries has important consequences for both classical and quantum dynamics in these backgrounds~\cite{desousa1990,OA,HH}. A thorough investigation of null geodesics in the spiral dislocation background is therefore essential from multiple standpoints. Theoretically, it provides a deeper understanding of massless field propagation in spacetimes with non-trivial topology and serves as a foundation for analyzing classical field equations, and particle dynamics in such geometries~\cite{vilenkin1994}. Phenomenologically, it enables predictions of light trajectories and lensing signatures in analog models of gravity, particularly those realized in structured media such as metamaterials and engineered photonic systems~\cite{Leonhardt2009, chen2010, leonhardt2006, smolyaninov2010}. Methodologically, an analytic treatment of geodesic equations facilitates the identification of hidden symmetries, effective potentials, and critical parameters that govern stability and phase space structure~\cite{crispino2008, hackmann2008}. Despite the rich structure of spiral dislocation geometries, most existing studies have focused on quantum mechanical treatments~\cite{OA,HH,New-2}, often neglecting the classical or optical interpretations of photon motion. This limits their applicability to broader contexts, such as numerical simulations, laboratory analogs, and pedagogical models.

\vspace{0.10cm}
\setlength{\parindent}{0pt}

In this work, we present a comprehensive analysis of null geodesics in the (2+1)-dimensional sector of the spiral dislocation spacetime, excluding the contribution of conserved linear momentum along the symmetry axis~\cite{OA,HH}. We begin by establishing the spacetime metric and identifying the associated Killing symmetries that yield conserved quantities. The geodesic equations are derived via the Lagrangian formalism and analyzed using the effective potential approach. We systematically quantify the influence of the dislocation parameter \(\beta\) on photon propagation, with explicit analytic derivations for the angular trajectory and the effective optical behavior. Particular attention is devoted to the limiting case \(\beta \to 0\), which recovers the standard Minkowski geometry, providing a benchmark for comparison. Beyond geometric optic analysis, we extend our investigation to wave optics \cite{semra-2025,Rop2}, determining the space- and frequency-dependent refractive index and exploring its physical implications. These results elucidate the connection between the geometry of torsional defects and their observable signatures in both gravitational and condensed matter systems.

\vspace{0.10cm}
\setlength{\parindent}{0pt}

The structure of this manuscript is as follows. Section~\ref{Sec:2} introduces the (2+1)-dimensional spiral dislocation metric and examines its geometric properties. In Section~\ref{Sec:3}, we derive the null geodesic equations and analyze the associated conserved quantities. Section~\ref{Sec:4} discusses the physical implications of photon motion within the dislocated background, focusing on the role of the effective potential. Section~\ref{Sec:5} characterizes the angular behavior of the photon trajectories. Section~\ref{Sec:7} extends the analysis to wave optics, deriving the refractive index and the optical response of the medium. Finally, Section~\ref{Sec:8} summarizes the main conclusions and outlines directions for future research.

\begin{figure}[ht]
\centering
\includegraphics[scale=0.5]{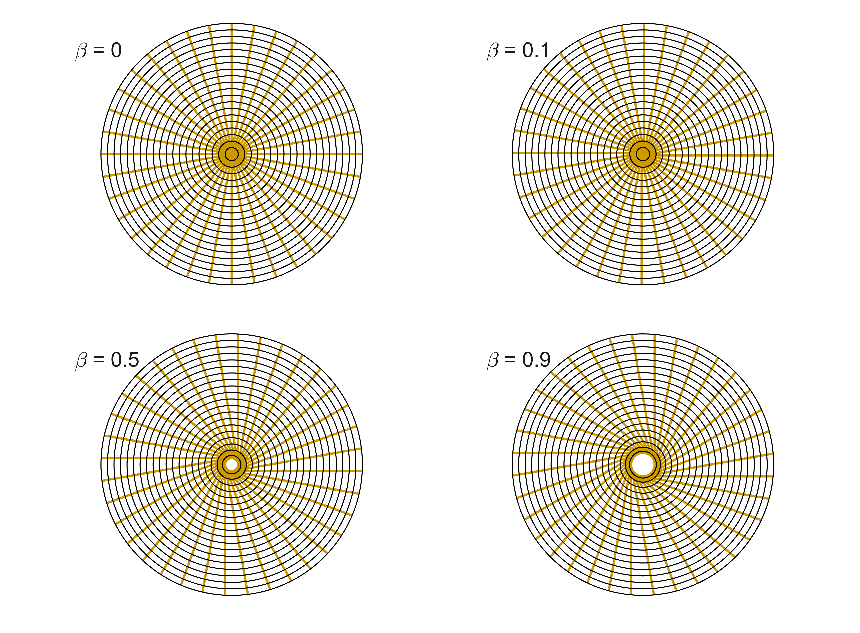}\\
\caption{\fontsize{7.7}{8.7}\selectfont Visualization of the spatial geometry induced by a spiral dislocation for various values of the torsion parameter $\beta$. Each subplot represents the deformed polar grid structure in the presence of a screw-type topological defect, where the usual polar coordinates $(r, \phi)$ are transformed into Cartesian coordinates $(x, y)$ according to the relations~\cite{HH}:
\(x(r, \phi) = r \cos\phi + \beta \sin\phi, \, y(r, \phi) = r \sin\phi - \beta \cos\phi\).
The parameter $\beta$ quantifies the strength of the intrinsic torsion of the spatial manifold. For $\beta = 0$, the geometry reduces to the standard flat polar plane, while increasing $\beta$ introduces increasing torsional distortion, as seen in the shifting and skewing of the radial and circular lines. This geometric effect arises from the non-diagonal metric component $g_{r\phi} = \beta$ in the spatial line element:
 \(d\tilde{s}^2 = dr^2 + 2 \beta\, dr\, d\phi + (\beta^2 + r^2)\, d\phi^2\).
}
\label{fig:SD}
\end{figure}

\section{Metric and Geometric Setup} \label{Sec:2}

In this section, we introduce the (2+1)-dimensional spacetime structure generated by a spiral dislocation, a type of topological defect that induces a helical distortion in the underlying spatial manifold. The geometry associated with this defect is encapsulated by the following line element~\cite{OA,HH}:
\begin{equation}
ds^2 = -dt^2 + dr^2 + 2\, \beta \, dr \, d\phi + (\beta^2 + r^2)\, d\phi^2,
\label{metric}
\end{equation}
where \( t \in (-\infty, \infty) \) denotes the temporal coordinate, and \( (r, \phi) \) represent the radial and azimuthal coordinates in the spatial section orthogonal to time. The parameter \( \beta \) quantifies the strength of the dislocation, effectively controlling the magnitude of torsional distortion induced by the defect~\cite{torsion}. The spacetime described by Eq.~\eqref{metric} deviates from a flat cylindrical geometry through the presence of a non-diagonal metric component \( g_{r\phi} = \beta \), which introduces coupling between the radial and angular directions. Additionally, the angular part of the metric is modified to \( g_{\phi\phi} = \beta^2 + r^2 \), reflecting the presence of intrinsic torsion. This feature leads to locally anisotropic spatial intervals and can be interpreted as the imprint of a screw-type dislocation, a structure familiar from dislocation theory in condensed matter physics and certain gravitational analogues~\cite{vilenkin1994,mermin1979,torsion,furtado1994,katanaev1992}. As visualized in Figure~\ref{fig:SD}, these deformations manifest in nontrivial spatial holonomies and anisotropies. The off-diagonal term \( g_{r\phi} \) breaks the orthogonality of the coordinate basis vectors, and constant-time hypersurfaces acquire an intrinsic twist, highlighting the topological influence of the defect.

\section{Lagrangian and conserved quantities} \label{Sec:3}

To study the propagation of massless particles, such as photons, in this spiral dislocation background~\eqref{metric}, we consider null geodesics characterized by the condition \( ds^2 = 0 \). Substituting this constraint into the metric~\eqref{metric} yields the following equation governing the geodesic motion~\cite{Rop1,semra-2025,Rop2}:
\begin{equation}
- \dot{t}^2 + \dot{r}^2 + 2\, \beta\, \dot{r}\, \dot{\phi} + (\beta^2 + r^2)\, \dot{\phi}^2 = 0,
\label{null_condition}
\end{equation}
where the overdot denotes differentiation with respect to an affine parameter \( \lambda \) that parametrizes the photon trajectory. The corresponding Lagrangian for the geodesic motion~\cite{Rop1,semra-2025,Rop2} in this background reads:
\begin{equation}
2\,\mathcal{L} = g_{\mu\nu} \dot{x}^\mu \dot{x}^\nu =  -\dot{t}^2 + \dot{r}^2 + 2\, \beta\, \dot{r} \dot{\phi} + (\beta^2 + r^2) \dot{\phi}^2,\quad  \mu, \nu \in \{ t, r, \phi \}.
\label{Lagrangian}
\end{equation}
For null geodesics, the condition \( \mathcal{L} = 0 \) holds~\cite{Rop1,semra-2025,Rop2}. The spacetime is both stationary and axisymmetric, as the metric components are independent of the coordinates \( t \) and \( \phi \). These symmetries imply the existence of two Killing vector fields, \( \partial_t \) and \( \partial_\phi \), which in turn ensure the conservation of the canonical momenta conjugate to these coordinates. The canonical momenta are defined via \cite{Landau}:
\begin{equation}
p_\mu = \frac{\partial \mathcal{L}}{\partial \dot{x}^\mu} = g_{\mu\nu} \dot{x}^\nu.
\label{canonical_momenta_def}
\end{equation}
Using the metric components from Eq.~\eqref{metric}, we compute:
\begin{align}
p_t &= \frac{\partial \mathcal{L}}{\partial \dot{t}} = g_{tt}\, \dot{t} = -\dot{t}, \nonumber\\
p_\phi &= \frac{\partial \mathcal{L}}{\partial \dot{\phi}} = g_{\phi r}\, \dot{r} + g_{\phi \phi}\, \dot{\phi} = \beta\, \dot{r} + (\beta^2 + r^2)\, \dot{\phi}.
\label{canonical_momenta}
\end{align}
We define the conserved quantities associated with the Killing vectors as:
\begin{equation}
p_t = -\mathcal{E}, \qquad p_\phi = \ell,
\label{conserved_quantities}
\end{equation}
where \( \mathcal{E} > 0 \) denotes the energy of the photon and \( \ell \) represents its angular momentum. These constants of motion remain invariant along null geodesics and play a pivotal role in simplifying the equations of motion, ultimately enabling analytical or numerical integration of the photon trajectories.

\section{Radial Dynamics and Effective Potential} \label{Sec:4}

\begin{figure}[ht]
\centering
\includegraphics[scale=0.5]{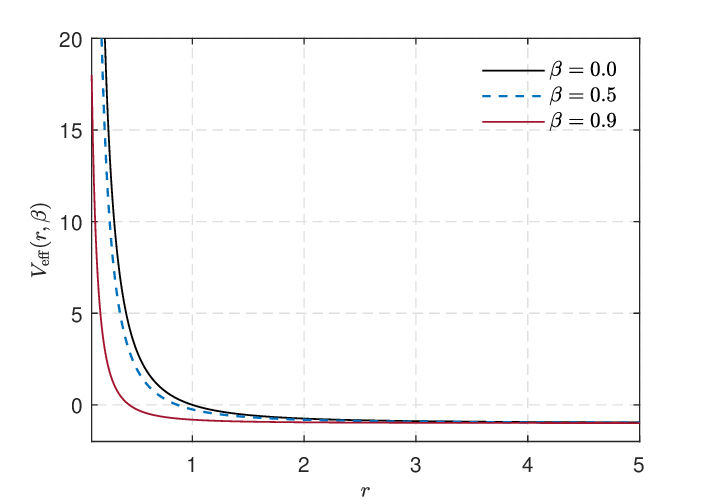}\\
\caption{\fontsize{7.7}{8.7}\selectfont Effective potential $V_{\rm eff}(r)$ plotted as a function of the radial coordinate $r$ for multiple values of the spiral dislocation parameter $\beta$. The plot clearly illustrates how increasing $\beta$ modifies the potential profile, revealing the influence of the dislocation parameter on the radial behavior of the system. Fixed parameters are energy $\mathcal{E} = 1$ and angular momentum $\ell = 1$. The curves emphasize the distinct potential landscapes for each $\beta$, highlighting key features relevant to the dynamics of null trajectories in the given geometry.
}
\label{fig:eff-pot-1}
\end{figure}

By inverting \eqref{canonical_momenta}, we express the coordinate velocities \(\dot{t}\) and \(\dot{\phi}\) in terms of \(\dot{r}\), \(\mathcal{E}\), and \(\ell\):
\begin{equation}
\dot{t} = \mathcal{E}, \quad
\dot{\phi} = \frac{\ell - \beta\, \dot{r}}{\beta^2 + r^2}.
\label{velocities}
\end{equation}
Substituting these into the null condition \eqref{null_condition} results in a quadratic equation for \(\dot{r}\):
\begin{equation}
- \mathcal{E}^2 + \dot{r}^2 + 2\, \beta\, \dot{r}\, \frac{\ell - \beta\, \dot{r}}{\beta^2 + r^2} + (\beta^2 + r^2) \left( \frac{\ell - \beta\, \dot{r}}{\beta^2 + r^2} \right)^2 = 0.
\end{equation}
Expanding and simplifying, a remarkable cancellation of the linear terms in \(\dot{r}\) occurs, yielding
\begin{equation}
- \mathcal{E}^2 + \dot{r}^2 \frac{r^2}{\beta^2 + r^2} + \frac{\ell^2}{\beta^2 + r^2} = 0.
\end{equation}
Rearranging, we isolate \(\dot{r}^2\) as
\begin{equation}
\dot{r}^2 = \frac{\mathcal{E}^2 (\beta^2 + r^2) - \ell^2}{r^2}.
\label{radial_velocity}
\end{equation}
Equation \eqref{radial_velocity} governs the radial motion of photons within the spiral dislocation geometry. Interpreting this equation in analogy to classical mechanics, it can be rewritten in terms of an effective potential \(V_{\rm eff}(r)\):
\begin{equation}
\dot{r}^2 + V_{\rm eff}(r) = 0,
\label{effective_potential_eq}
\end{equation}
where
\begin{equation}
V_{\rm eff}(r) = -\frac{\mathcal{E}^2 (\beta^2 + r^2) - \ell^2}{r^2} = -\mathcal{E}^2 - \frac{\mathcal{E}^2 \beta^2 - \ell^2}{r^2}.
\label{effective_potential}
\end{equation}
This effective potential (see Figure \ref{fig:eff-pot-1}) encapsulates the combined effects of the photon's energy, angular momentum, and the geometric parameter \(\beta\) on radial propagation. The term proportional to \(1/r^2\) acts analogously to a centrifugal barrier modified by the dislocation parameter, significantly affecting photon trajectories near the defect core.

\section{Angular Trajectories and Their Properties} \label{Sec:5}

In this section, we investigate the angular evolution of a photon traversing a spacetime endowed with a spiral dislocation, a geometric defect characterized by torsion in the absence of curvature. Unlike standard Riemannian spacetimes, where the connection is Levi-Civita and fully determined by the metric, the presence of torsion modifies the affine structure of the manifold, yielding a geometry where parallel transport is sensitive to topological defects. In such a setting, even massless particles such as photons experience trajectory deviations, not through curvature-induced geodesic bending but via torsion-mediated geometric shifts. Our aim is to derive the exact angular trajectory \(\phi(r)\) of a light-ray propagating through this torsion-dominated spacetime and to provide a detailed interpretation of the resulting expression. The dynamics of the photons are encapsulated in the first-order differential equation governing its angular coordinate as a function of radial position \(r\):
\begin{equation}
\frac{d\phi}{dr} = \frac{\dot{\phi}}{\dot{r}}=\frac{\ell}{(\beta^2 + r^2)\, \dot{r}} - \frac{\beta}{\beta^2 + r^2}.
\label{eq:dphi_dr}
\end{equation}
The first term corresponds to the standard angular evolution from conservation laws, modified by the torsion-induced change in the effective radial metric. The second term, however, is a direct geometric contribution arising purely from torsion and independent of the photon’s dynamical quantities $(\mathcal{E},\ell)$, representing a geometric phase-like effect induced by the torsional defect. To proceed, we invoke the null condition \(ds^2 = 0\) for the photon's worldline, which leads to the following expression for the radial velocity:
\begin{equation}
\dot{r} = \frac{1}{r}\,\sqrt{\mathcal{E}^2 (\beta^2 + r^2) - \ell^2}.
\label{eq:rdot}
\end{equation}
Substituting Eq.~\eqref{eq:rdot} into Eq.~\eqref{eq:dphi_dr}, and introducing the impact parameter \(b = \ell/\mathcal{E}\), we obtain:
\begin{equation}
\frac{d\phi}{dr} = \frac{b\, r}{(\beta^2 + r^2)\, \sqrt{\beta^2 + r^2 - b^2}} - \frac{\beta}{\beta^2 + r^2}.
\label{eq:dphi_dr_b}
\end{equation}
To compute the net angular displacement as the photon moves from the source position \(r = r_s\) to an arbitrary radial position \(r\), we integrate:
\begin{equation}
\phi(r) - \phi(r_s) = \int_{r_s}^{r} \left[ \frac{b\, r'}{(\beta^2 + r'^2)\, \sqrt{\beta^2 + r'^2 - b^2}} - \frac{\beta}{\beta^2 + r'^2} \right] dr',
\label{eq:phi_integral}
\end{equation}
where \(r'\) is the integration variable over radial distance. We separate Eq.~\eqref{eq:phi_integral} into two integrals for clarity:
\begin{align}
I_1 &= b \int_{r_s}^{r} \frac{r'}{(\beta^2 + r'^2)\, \sqrt{\beta^2 + r'^2 - b^2}}\, dr', \label{eq:I1} \\
I_2 &= \beta \int_{r_s}^{r} \frac{1}{\beta^2 + r'^2}\, dr'. \label{eq:I2}
\end{align}
The second integral is elementary \cite{Abramowitz}:
\begin{equation}
I_2 = \arctan\left( r/\beta \right) - \arctan \left( r_s/\beta \right).
\label{eq:I2_result}
\end{equation}
To evaluate \(I_1\), we perform the substitution \(x = \beta^2 + r'^2\), so that \(dx = 2r' \, dr'\), transforming Eq.~\eqref{eq:I1} into:
\begin{equation}
I_1 = \frac{b}{2} \int_{x_s}^{x} \frac{1}{x \sqrt{x - b^2}} \, dx,
\label{eq:I1_x}
\end{equation}
where \(x_s = \beta^2 + r_s^2\) and \(x = \beta^2 + r^2\). Introducing \(y = \sqrt{x - b^2}\), so \(x = y^2 + b^2\) and \(dx = 2y\, dy\), yields:
\begin{equation}
I_1 = b \int_{y_s}^{y} \frac{1}{y^2 + b^2}\, dy,
\label{eq:I1_y}
\end{equation}
with
\[
y = \sqrt{\beta^2 + r^2 - b^2}, \quad y_s = \sqrt{\beta^2 + r_s^2 - b^2}.
\]
This integral evaluates to \cite{Abramowitz}:
\begin{equation}
I_1 = \arctan\left( \frac{\sqrt{\beta^2 + r^2 - b^2}}{b} \right) - \arctan\left( \frac{\sqrt{\beta^2 + r_s^2 - b^2}}{b} \right).
\label{eq:I1_result}
\end{equation}
Thus, the total angular displacement between the source at \(r_s\) and the point \(r\) is:
\begin{equation}
\begin{split}
\phi(r) - \phi(r_s) =\ & \arctan\left( \frac{\sqrt{\beta^2 + r^2 - b^2}}{b} \right) - \arctan\left( \frac{\sqrt{\beta^2 + r_s^2 - b^2}}{b} \right) \\
& - \left[ \arctan\left( \frac{r}{\beta} \right) - \arctan\left( \frac{r_s}{\beta} \right) \right].
\end{split}
\label{eq:phi_final}
\end{equation}

\vspace{0.10cm}
\setlength{\parindent}{0pt}

To interpret the physical deflection angle, we consider a source located at radial position \(r_s\) and an observer positioned at radial position \(r_o\), both taken sufficiently far from the torsional defect such that the spacetime approaches a flat but topologically nontrivial limit. The measurable deflection angle \(\Delta \phi\) experienced by a photon emitted from the source and detected at the observer is given by the total angular change between these two points \cite{ovgun}:
\begin{equation}
\Delta \phi = \phi(r_o) - \phi(r_s).
\label{eq:deflection_angle}
\end{equation}
Here, \(r_s\) and \(r_o\) set the physical boundaries of the photon's trajectory, and the integration variable \(r\) runs over the radial positions between them. The impact parameter \(b\) determines the minimal approach distance of the photon to the torsional defect and encodes the initial conditions of the photon's motion. Because the angular displacement formula contains both a dynamical term (dependent on \(b\)) and a purely geometric torsion-induced term (independent of the photon's energy or angular momentum), the net deflection angle \(\Delta \phi\) is influenced by both physical and geometric factors. This clear identification of source and observer locations completes the physical picture: the deflection angle is not merely a theoretical artifact but corresponds to a measurable angular shift in the photon's path between two spatially separated points, analogous to standard gravitational lensing but here induced by torsion rather than curvature.

\begin{figure}[ht]
\centering
\includegraphics[scale=0.50]{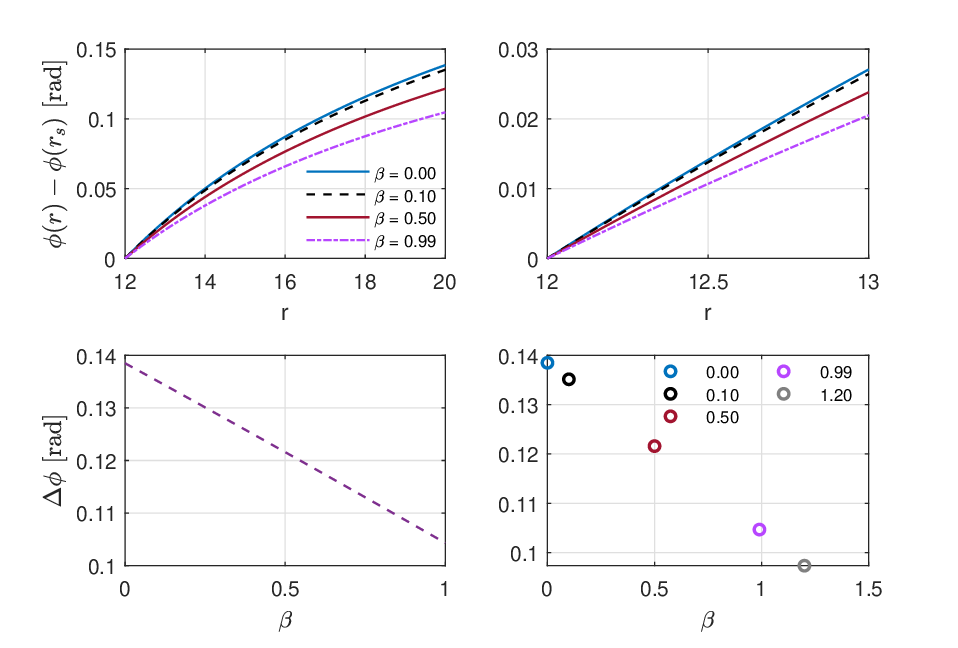}\\
\caption{\fontsize{7.7}{8.7}\selectfont The impact of the parameter $\beta$ on photon angular trajectories and the net deflection angle is illustrated over the radial domain $r \in [r_s, r_o]$ with source radius $r_s = 12$, observer radius $r_o = 20$, and impact parameter $b = 4$. The first subplot shows the angular displacement $\phi(r) - \phi(r_s)$ for selected values of $\beta = 0, 0.1, 0.5, 0.99$, revealing how increasing $\beta$ progressively decreases the bending of photon trajectories. The second subplot provides a zoomed-in view near the source radius $r_s$, highlighting subtle differences in angular displacement where the photon path is most sensitive to changes in $\beta$. The third subplot presents the continuous variation of the net deflection angle $\Delta \phi = \phi(r_o) - \phi(r_s)$ as a function of $\beta$ in the range $[0,1]$, showing a smooth decrease in deflection magnitude and a zero-crossing at approximately $\beta \approx 0.658$. This zero-crossing marks a critical point where the photon experiences no net bending, signaling a transition between effective attractive and repulsive regimes. The fourth subplot depicts discrete deflection angle values at the selected $\beta$ points, emphasizing the physically relevant regimes where the deflection changes sign. Together, these results underscore the pivotal role of $\beta$ in controlling both the magnitude and direction of photon deflection, with important implications for understanding lensing and scattering phenomena induced by torsion in the underlying geometry. The combined use of full-range plots and zoomed insets allows for detailed examination of the photon trajectory behavior and the identification of critical parameter values that govern the transition between bending regimes.}
\label{fig:1}
\end{figure}

\vspace{0.10cm}
\setlength{\parindent}{0pt}

Moreover, this torsion-induced deflection can be realized and tested in laboratory analog gravity systems. In these platforms, such as photonic crystals, metamaterials, or other engineered media, effective torsion-like geometric defects can be created by introducing spiral dislocations or similar topological structures. Photons or quasiparticles propagating through these media acquire phase shifts and angular deviations akin to those computed in the gravitational analog. In laboratory setups, the source and observer correspond to well-defined positions of photon emitters and detectors embedded within or around the medium. The parameter \(\beta\) models the strength or pitch of the engineered torsion defect. Measuring angular deflections or interference patterns in such controlled experiments provides a unique window into torsion physics, otherwise difficult to observe directly in astrophysical or cosmological settings. Hence, the exact angular evolution formula derived here not only describes fundamental photon trajectories in torsional spacetimes but also serves as a theoretical foundation for experimental explorations of torsion-induced optical phenomena in laboratory analogs.

\vspace{0.10cm}
\setlength{\parindent}{0pt}

In the limiting case where torsion vanishes, \(\beta \to 0\), the angular displacement reduces to the flat spacetime result:
\begin{equation}
\phi(r) - \phi(r_s) = \arctan\left( \frac{\sqrt{r^2 - b^2}}{b} \right) - \arctan\left( \frac{\sqrt{r_s^2 - b^2}}{b} \right),
\end{equation}
recovering the expected standard behavior without geometric phase contributions. The corresponding angular displacements for various values of $\beta$ are shown in Figure~\ref{fig:1}. Consequently, the angular evolution of a photon in a spacetime with spiral dislocation comprises a superposition of dynamical and torsion-induced geometric contributions, with the deflection angle explicitly determined by the radial positions of the source and observer. This framework connects gravitational theory with experimentally accessible laboratory analogs, offering new avenues for investigating torsion physics through photonic systems.

\vspace{0.10cm}
\setlength{\parindent}{0pt}

Physically, our exact solutions reveal that the torsion-induced correction decreases the total angular span covered by the photon compared to the torsion-free scenario. Unlike classical gravitational lensing caused by positive curvature sourced by mass-energy, which increases the bending angle, here torsion acts as a geometric twist that effectively ``unwinds" the photon trajectory. This results in a negative contribution to the angular deviation, demonstrating that topological defects characterized by torsion produce optical effects fundamentally distinct from those arising from Riemannian curvature. The modification arises because the substitution \(r^{2} \to \beta^{2} + r^{2}\) alters the effective radial potential governing photon dynamics, shifting the turning points and changing the curvature of null geodesics. Simultaneously, the angular evolution equation acquires explicit \(\beta\)-dependent terms both in the denominator and as an additive correction, reflecting the screw-like geometric structure of the medium. Importantly, the absence of Riemann curvature in this model confirms that the observed angular deflection is a pure torsional effect, exemplifying how non-Riemannian geometric structures imprint measurable signatures on light propagation even in flat spacetime. This analytic framework thus provides a rare and valuable example where deflection of light in a torsion-dominated geometry can be computed, yielding clear insight into the influence of topological defects on photon trajectories. Beyond its fundamental theoretical interest in alternative gravity and geometric theories, this effect has concrete analogs in condensed matter systems where screw dislocations and torsional defects critically affect electronic and photonic transport properties. The results presented here offer a rigorous foundation for exploring such phenomena and underscore the rich relationship between geometry, topology, and photon propagation in environments endowed with torsion.

\vspace{0.10cm}
\setlength{\parindent}{0pt}

After analyzing the behavior of light using the geometric optics approach in spiral dislocation spacetime, we now shift our focus to the wave optics regime. This shift enables us to account for diffraction and interference effects that are beyond the scope of ray-based methods. Investigating wave propagation within the same spacetime framework offers a more complete picture of how torsion-induced geometry affects optical phenomena.

\section{Wave optics} \label{Sec:7}

\vspace{0.10cm}
\setlength{\parindent}{0pt}

In this section, we analyze wave optics in the spiral dislocation background and examine the optical response of the two-dimensional manifold equipped with the corresponding spatial metric:
\begin{equation}
d\tilde{s}^{2} = dr^{2} + 2 \beta\, dr\, d\phi + \left( r^{2} + \beta^{2} \right) d\phi^{2},
\label{eq:metric}
\end{equation}
where the radial coordinate satisfies $r \geq 0$ and the angular coordinate $\phi \in [0,2\pi)$ parametrizes a circle. The associated metric tensor and its determinant are given by
\begin{equation}
g_{ij} =
\begin{pmatrix}
1 & \beta \\[6pt]
\beta & r^{2} + \beta^{2}
\end{pmatrix}, \quad
\det(g) = r^{2},\quad  i, j \in \{ r, \phi \},
\label{eq:metric_tensor}
\end{equation}
indicating the absence of intrinsic curvature but the presence of anisotropy. The inverse metric reads
\begin{equation}
g^{ij} = \frac{1}{r^{2}}
\begin{pmatrix}
r^{2} + \beta^{2} & -\beta \\[6pt]
-\beta & 1
\end{pmatrix}.
\label{eq:inverse_metric}
\end{equation}
The scalar wave dynamics in such a background are governed by the Helmholtz equation, $\Delta_{g} \Psi + k^2 \Psi = 0$, where $\Delta_{g}$ is the Laplace-Beltrami operator associated with the metric~\cite{semra-2025,Rop2}, and $k$ denotes the wave number, related to the energy or angular frequency of the wave $(k=\omega/c)$. In explicit form, $\Delta_{g} \Psi$ is given by
\begin{equation}
\Delta_{g} \Psi = \frac{1}{\sqrt{|g|}} \partial_{i} \left( \sqrt{|g|} \, g^{ij} \partial_{j} \Psi \right),
\label{eq:laplace_beltrami_general}
\end{equation}
which in these coordinates takes the following form:
\begin{equation}
\begin{split}
\Delta_{g} \Psi = \frac{1}{r} \partial_{r} \left[ r \left( g^{rr} \partial_{r} \Psi + g^{r\phi} \partial_{\phi} \Psi \right) \right] + \partial_{\phi} \left( g^{\phi r} \partial_{r} \Psi + g^{\phi \phi} \partial_{\phi} \Psi \right).
\label{eq:laplace_beltrami_explicit}
\end{split}
\end{equation}
Assuming separable solutions of the form \(\Psi(r, \phi) = \psi(r)\, e^{i\, m\, \phi}, \, m \in \mathbb{Z}\), and substituting into the Helmholtz equation, $\Delta_{g} \Psi + k^2 \Psi = 0$ \cite{Rop2}, leads to the radial differential equation \cite{OA}:
\begin{equation}
\left(1+\frac{\beta^2}{r^2}\right) \psi''+\left(\frac{1}{r}-\frac{\beta^2}{r^3}-\frac{2im\beta}{r^2}\right)\psi'+\left(\frac{im\beta}{r^3}-\frac{m^2}{r^2}+k^2\right)\psi=0.
\end{equation}
Here, \( \psi' \) denotes the derivative of \( \psi \) with respect to the independent variable. In this equation, the imaginary term $-2 i \beta m/r^2\,  \psi'(r)$ arises from the off-diagonal component $g_{r\phi}$ and acts as a geometric gauge coupling \(A_r \sim \frac{m\beta}{r^2}\), entangling the effects of the magnetic quantum number \(m\) with radial propagation via an effective gauge connection. This also means that $\beta$ acts as a topological (geometric) charge inducing a gauge-like coupling. Now, let us normalize the coefficient of $\psi''(r)$ and divide by $\left(1 + \beta^2/r^2\right)$ to obtain
\begin{equation}
\psi''(r) + p(r)\, \psi'(r) + q(r)\, \psi(r) = 0,
\end{equation}
where
\begin{align}
p(r) = \frac{r - \frac{\beta^2}{r} - 2i m \beta}{r^2 + \beta^2}, \quad q(r) = \frac{k^2 r^2 - m^2 + \frac{i m \beta}{r}}{r^2 + \beta^2}.
\end{align}
To eliminate the first derivative term ($\propto \psi'$), we use the standard transformation:
\begin{equation}
\psi(r) = \mu(r)\, \tilde{\psi}(r), \qquad \mu(r) = \exp\left( -\frac{1}{2} \int P(r)\, dr \right).
\end{equation}
This yields a one-dimensional Schrödinger-like equation for $\tilde{\psi}(r)$:
\begin{equation}
\tilde{\psi}''(r) + \left[ q(r) - \frac{1}{2} p'(r) - \frac{1}{4} p(r)^2 \right] \tilde{\psi}(r) = 0.\label{WE-E}
\end{equation}
Which can be simplified, in a straightforward manner, to read
\begin{equation}
\tilde{\psi}''(r)+\mathcal{V}(r,k)\,\tilde{\psi}(r)=0,\label{S-F}
\end{equation}
where
\begin{equation}
\mathcal{V}(r,k)=\frac{k^2r^4+(k^2\beta^2-m^2+1/4)\,r^2-3\beta^2/2-3\beta^4/(4r^2)}{(r^2+\beta^2)^2}.\label{HH1}
\end{equation}
Moreover, upon the change of variables $x=r^2+\beta^2\Rightarrow r=\sqrt{x-\beta^2}$, to obtain Bessel like equation that admits a solution in the form of
\begin{equation}
\begin{split}
&\tilde{\psi}(x) = \frac{\sqrt{x}}{(x-\beta^2)^{1/4}} \left( \mathcal{C}_1 \, J_m(k \sqrt{x}) + \mathcal{C}_2 \, Y_m(k \sqrt{x}) \right)\implies \\
&\tilde{\psi}(r) = \sqrt{r + \frac{\beta^2}{r}} \left( \mathcal{C}_1 \, J_m \bigl(k \sqrt{r^2 + \beta^2}\bigr) + \mathcal{C}_2 \, Y_m \bigl(k \sqrt{r^2 + \beta^2}\bigr) \right),
\end{split}
\end{equation}
where $J_m$ and $Y_m$ are Bessel functions of the first and second kinds, respectively. In particular, when the dislocation parameter $\beta\to 0$ our $\tilde{\psi}(r)|_{r\to 0}\to 0$. Moreover, at $\beta=0$ our $\mathcal{V}(r,k)$ in (\ref{HH1}) simplifies to read $\mathcal{V}(r,k)=k^2-(m^2-1/4)/r^2$ and eventually Eq.(\ref{S-F}) admits a solution $\tilde{\psi}(r)=\sqrt{r}\left(\mathcal{C}_1 \,J_m(k{r})+\mathcal{C}_2\,Y_m(kr)\right)$. In this case, however, $\psi(r)=\tilde{\psi}(r)/\sqrt{r}=\mathcal{C}_1 \,J_m(k{r})+\mathcal{C}_2\,Y_m(kr)$ which should be at least finite at the origin $r=0$ and consequently $\mathcal{C}_2=0$ to eventually obtain $\psi(r)=\mathcal{C}_1\,J_m(kr)$ as the exact solution to the problem at hand.

\vspace{0.10cm}
\setlength{\parindent}{0pt}

To interpret this setup using an optical analogy, we introduce a position- and frequency-dependent refractive index $n(r, k)$ by casting Eq.~\eqref{S-F} into the generalized Helmholtz-like form \cite{semra-2025}:
\begin{equation}
\tilde{\psi}''(r) + k^{2}\, n^{2}(r,k)\, \tilde{\psi}(r) = 0.
\label{eq:helmholtz_form}
\end{equation}
From which it follows that $n^2(r, k)=k^{-2}\,\mathcal{V}(r,k)$:
\begin{equation}
n^2(r, k) = \frac{
r^4 + \left( \beta^2 - \frac{m^2}{k^2} + \frac{1}{4k^2} \right) r^2 - 3\beta^2/(2k^2) - 3\beta^4/(4k^2 r^2)
}{(r^2 + \beta^2)^2}.\label{eq:n_squared_def}
\end{equation}
As the radial coordinate \(r\) tends to infinity, the inverse power terms vanish and \(n^{2}(r,k)\) asymptotically approaches unity. This indicates that waves propagating far from the dislocation core experience no effective medium modification and behave as if propagating in a free-like space with a refractive index \(n = 1\). Therefore, the influence of topological and geometric effects becomes negligible at large distances. Near the core, where \(r \to 0\), the inverse powers of \(r\) dominate the refractive index profile. Among these, the most singular contribution arises from the term
\begin{equation*}
- \frac{3\,\beta^4}{4\,k^2\, r^2\, (r^2 + \beta^2)^2},
\end{equation*}
which originates from the last term in Eq.~\eqref{eq:n_squared_def}. This negative definite term becomes extremely large in magnitude for small \(r\), driving \(n^2(r,k)\) to negative values near the origin. In this limit, the refractive index behaves as
\begin{equation*}
n(r,k) \sim i\,\sqrt{\frac{3\,\beta^4}{4\,k^2\, r^2\, (r^2 + \beta^2)^2}},
\end{equation*}
corresponding to a purely imaginary refractive index characteristic of a lossy medium. This implies exponential attenuation of electromagnetic waves, especially at low frequencies (\(k \to 0\)), leading to the emergence of a localized evanescent region where wave propagation is suppressed. Additional contributions also play a role. The term \((m^2 - 1/4)/(k^2 r^2)\) represents the centrifugal barrier, acting repulsively to push the wave function away from the origin. In contrast, the geometric term \(\beta^2/r^2\) partially offsets this effect, modifying the overall index profile. The combination of attractive and repulsive terms creates a complex refractive index structure near the core, enabling phenomena such as wave trapping or strong scattering. A critical condition for wave propagation is the sign of \(n^2(r,k)\). When \(n^2(r,k) > 0\), the refractive index is real and the waves can propagate. When \(n^2(r,k) < 0\), the refractive index becomes imaginary, leading to evanescent non-propagating modes. The strong singular term near \(r = 0\) makes it likely that \(n^2(r,k)\) becomes negative in this region, causing a strong localization. The frequency dependence introduced by the wavevector \(k\) is crucial. Since negative contributions scale as \(1/k^2\), they dominate at low frequencies, increasing the likelihood of \(n^2(r,k) < 0\) and thus suppressing propagation. At high frequencies, these corrections diminish and \(n^2(r,k) \to 1\), restoring free-wave behavior. This makes the medium act as a frequency-selective filter: low-frequency waves are suppressed, while high-frequency waves propagate nearly unaffected. In the special case where the dislocation parameter \(\beta\) vanishes, the refractive index reduces to \(n^2(r,k) = 1 -(m^2 - 1/4)/(k^2 r^2)\), which describes the standard scenario without geometric or topological defects. In the geometric optics limit \(k \to \infty\), the refractive index approaches \(n = 1\). The absence of \(\beta\) eliminates geometric corrections, isolating the contribution of the centrifugal barrier. This limiting case highlights the role of \(\beta\) in the introduction of higher-order geometric effects, as summarized in Table \ref{tab:refractive_index_limits}.

\begin{table}[!ht]
\centering
{\fontsize{7.4}{8.4}\selectfont
\renewcommand{\arraystretch}{1.3}
\setlength{\tabcolsep}{6pt}
\begin{tabular}{|>{\raggedright\arraybackslash}p{1cm}|>{\raggedright\arraybackslash}p{2.5cm}|>{\raggedright\arraybackslash}p{4cm}|}
\hline
Limit & \(n^{2}(r,k)\) behavior & Physical interpretation \\
\hline
\(r \to \infty\) &
\(n^{2} \to 1\) &
Free-space propagation; dislocation has no effect at large distances. \\
\hline
\(r \to 0\) &
Dominated by \(- \frac{3\beta^4}{4k^2 r^2 (r^2 + \beta^2)^2} \Rightarrow n^{2} \to -\infty\) &
Evanescent field near the core; strong wave localization and inhibited propagation due to geometric defect. \\
\hline
\(k \to 0\) &
Negative terms dominate; \(n^{2} \to -\infty\) &
Low-frequency waves are suppressed near the core; medium becomes opaque in this regime. \\
\hline
\(k \to \infty\) &
\(n^{2} \to 1\) &
High-frequency waves propagate freely; negligible influence from geometry or dislocation. \\
\hline
\(\beta \to 0\) &
\(n^{2} = 1 - \frac{m^{2} - \frac{1}{4}}{k^{2} r^{2}}\) &
Reduces to standard centrifugal barrier form; absence of topological or geometric effects. \\
\hline
\end{tabular}
}
\caption{\fontsize{7.7}{8.7}\selectfont Limiting behaviors of the refractive index \(n^{2}(r,k)\) and corresponding physical implications.}
\label{tab:refractive_index_limits}
\end{table}

\begin{figure}[ht]
\centering
\includegraphics[scale=0.60]{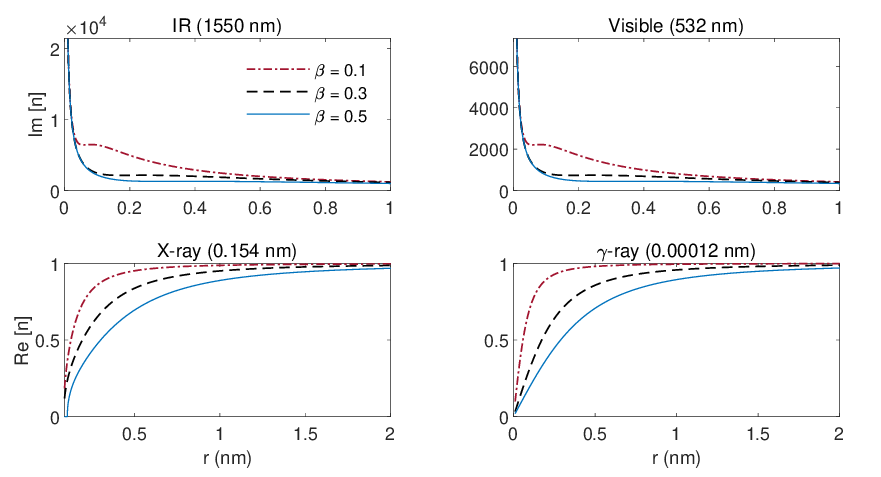}\\
\caption{\fontsize{7.7}{8.7}\selectfont Radial dependence of the refractive index \( n(r, k) \) for different electromagnetic regimes in the presence of a spiral dislocation. The imaginary part of \( n \) is shown for infrared (1550 nm) and visible (532 nm) wavelengths (top row), while the real part of \( n \) is plotted for X-ray (0.154 nm) and gamma-ray (0.00012 nm) wavelengths (bottom row). Each curve corresponds to a different dislocation parameter \( \beta = 0.1, 0.3, 0.5 \). The magnetic quantum number is fixed at \( m = 1 \). Increasing \( \beta \) modifies the local refractive structure, emphasizing the effect of topological defects on electromagnetic wave propagation in such a media. Here, we take $c = 3 \times 10^{17}\,\text{nm/s}$. All spatial parameters are expressed in nanometers.}
\label{fig:ref-index}
\end{figure}

\section{Summary and Conclusions} \label{Sec:8}

\vspace{0.10cm}
\setlength{\parindent}{0pt}

In this manuscript, we have presented a rigorous analytical study of null trajectories and scalar wave propagation in a (2+1)-dimensional spacetime featuring a spiral dislocation, a topological defect characterized by torsion in the absence of curvature. The dislocation parameter $\beta$ introduces a screw-type spatial deformation that modifies photon trajectories via the affine structure. Leveraging the axial and temporal symmetries of the background metric~\eqref{metric}, we identified conserved quantities associated with the Killing vectors $\partial_t$ and $\partial_\phi$, which reduce the geodesic equations to an effective one-dimensional radial form. The corresponding radial equation~\eqref{radial_velocity} encapsulates the combined effects of energy ($\mathcal{E}$), angular momentum ($\ell$), and torsion ($\propto \beta$). The effective potential~\eqref{effective_potential} reveals that torsion smooths the centrifugal barrier and enforces a finite turning point $r_{\text{min}} = \sqrt{b^2 - \beta^2} > 0$, where $b = \ell/\mathcal{E}$. This prohibits null trajectories from reaching the defect core, generating a classically forbidden region, $0\leq r<r_{\text{min}}$, which arises from torsion. Unlike curvature-induced singularities, this exclusion zone is a result of the spacetime topological structure. The angular trajectories derived from~\eqref{eq:dphi_dr_b} exhibit a torsion-dependent deflection term that decays as $1/r^2$ and contributes negatively to the total deflection angle, giving rise to distinct asymptotic photon behavior.

\vspace{0.10cm}
\setlength{\parindent}{0pt}

The analytic form of the angular trajectory $\phi(r)$ in~\eqref{eq:phi_final} shows that torsion alone produces ``observable" deviations in the photon paths. The modified geodesic equations produce an exact expression for the angular displacement, comprising a dynamical contribution dependent on the impact parameter $b$ and a geometric contribution proportional to $\beta$. The total deflection angle $\Delta \phi$ in Eq. \eqref{eq:deflection_angle} reflects the cumulative influence of these effects, with torsion reducing the angular span and effectively unwinding the trajectory. This contrasts with the curvature-induced bending in conventional gravitational lensing. The shift $r^2 \rightarrow r^2 + \beta^2$ in the radial potential alters the location of the turning point, yielding an exact expression for light deflection in a torsion-dominated, curvature-free spacetime, as given in~\eqref{eq:phi_final}. In the limit $\beta \to 0$, the standard flat spacetime result is recovered. These findings suggest experimental realization in analog systems such as photonic crystals or metamaterials capable of simulating torsional geometries. The framework thereby connects theoretical models of torsion to experimentally accessible optical phenomena, revealing a deep geometric-topological influence on photon propagation in non-Riemannian settings.

\vspace{0.10cm}
\setlength{\parindent}{0pt}

The analysis extends to scalar wave dynamics through a reformulation of the Helmholtz equation into a Schrödinger-like form in the torsional background. For optical interpretation, we introduce an effective refractive index $n(r,k)$ \eqref{eq:n_squared_def}, with $k = \omega/c$, leading to a generalized Helmholtz equation. As $r \to \infty$, the refractive index tends toward unity, indicating free-space propagation. Closer to the defect core, inverse power-law terms dominate the behavior of $n(r,k)$, particularly a strongly negative divergent term, resulting in wave suppression near the origin and defining an evanescent region. Additionally, the centrifugal term $-(m^2 - \frac{1}{4})/(k^2 r^2)$ contributes an angular momentum barrier. These components together shape a spatially varying refractive index that governs localization and attenuation of waves near the dislocation. Figure~\ref{fig:ref-index} shows the spatial profile of the complex refractive index $n(r, k)$ across distinct electromagnetic regimes. The imaginary part, governing attenuation, is presented for infrared ($\lambda = 1550\,\text{nm}$) and visible ($\lambda = 532\,\text{nm}$) wavelengths, while the real part, determining phase velocity, is shown for X-ray ($\lambda = 0.154\,\text{nm}$) and gamma-ray ($\lambda = 0.00012\,\text{nm}$) domains, where $\lambda=2\pi/k$. Variations in the torsion strength $\beta \in \{0.1, 0.3, 0.5\}$ significantly alter the refractive structure, emphasizing the role of topological defects in shaping electromagnetic mode behavior in structured media.

\vspace{0.10cm}
\setlength{\parindent}{0pt}

Our analysis shows that increasing the spiral dislocation parameter $\beta$ systematically modifies both null and scalar wave dynamics. Larger values of $\beta$ establish a finite torsion scale that regularizes the potential near the core, eliminating singular behavior and setting a minimum turning radius $r_{\min} = \sqrt{b^2 - \beta^2}$. This results in a topological barrier independent of curvature. As $\beta$ increases, null trajectories become progressively less curved, with a corresponding monotonic decrease in the deflection angle. Torsion thereby alters the effective inertial frame and acts as a tunable parameter controlling angular deviation. For scalar waves, higher $\beta$ adjusts the effective angular momentum contribution and enhances confinement through the torsion-modified refractive index. In both cases, torsion operates as a geometric regulator, dictating wave and geodesic behavior without invoking curvature effects.

\vspace{0.10cm}
\setlength{\parindent}{0pt}

In this study, we establish a foundational framework for light propagation in torsion-rich media, which holds significant implications for analog gravity, topological photonics, and metamaterial design. Future work may involve extensions to spinor and vector fields, incorporation of external electromagnetic interactions, and investigation of time-dependent or anisotropic torsional configurations. The results may highlight the important and often overlooked role of torsion in shaping the causal, optical, and spectral properties of waves in geometrically complex spacetimes.

\section*{CRediT authorship contribution statement}

\textbf{Semra Gurtas Dogan}: Conceptualization, Methodology, Investigation, Writing – Review and Editing, Formal Analysis, Validation, Visualization.\\
\textbf{Omar Mustafa}: Conceptualization, Methodology, Investigation, Writing – Review and Editing, Formal Analysis, Validation, Visualization.\\
\textbf{Abdullah Guvendi}: Conceptualization, Methodology, Investigation, Writing – Review and Editing, Formal Analysis, Validation, Visualization.\\
\textbf{Hassan Hassanabadi}: Conceptualization, Methodology, Investigation, Writing – Review and Editing, Formal Analysis, Validation, Visualization.\\

\section*{Data availability}

The authors confirm that the data supporting the findings of this study are available within the article.

\section*{Conflicts of interest statement}

The authors have disclosed no conflicts of interest.

\section*{Funding}

This research has not received any funding.

\end{document}